\documentclass[twocolumn]{aastex63}
\usepackage{amssymb,amsmath}
\usepackage{graphicx}
\usepackage{xcolor}
\usepackage{multirow}
\usepackage[T1]{fontenc}
\usepackage{ae,aecompl}
\usepackage{newtxtext,newtxmath}
\usepackage{longtable,listings}

\def\msig{$M_{\rm BH}-\sigma$\ }

\def\obj{SDSS J1241+2602}

\submitjournal{ApJ}

\shorttitle{Unobscured Type-1.9 AGN}
\shortauthors{Zhang XueGuang}

\begin{document}

\title{Unobscured central broad line regions in Type-1.9 AGN SDSS J1241+2602}

\correspondingauthor{XueGuang Zhang}%
\email{xgzhang@gxu.edu.cn}
\author{XueGuang Zhang$^{*}$}
\affiliation{Guangxi Key Laboratory for Relativistic Astrophysics, School of Physical Science and Technology,
GuangXi University, Nanning, 530004, P. R. China}

\begin{abstract} 
In this manuscript, strong evidence is reported to support unobscured broad line regions (BLRs) in Type-1.9 AGN SDSS J1241+2602 
with reliable broad H$\alpha$ but no broad H$\beta$. Commonly, disappearance of broad H$\beta$ can be explained by the AGN unified 
model expected heavily obscured BLRs in Type-1.9 AGN. Here, based on properties of two kinds of BH masses, the virial 
BH mass and the BH mass through the \msig relation, an independent method is proposed to test whether are there unobscured central 
BLRs in a Type-1.9 AGN. By the reliable measurement of stellar velocity dispersion about 110$\pm$12km/s through the host galaxy 
absorption features in \obj, the BH mass through the \msig relation is consistent with the virial BH mass 
$(3.43\pm1.25)\times10^7{\rm M_\odot}$ determined through properties of the observed broad H$\alpha$ without considering effects 
of obscurations in SDSS J1241+2602. Meanwhile, if considering heavily obscured BLRs in SDSS J1241+2602, the reddening corrected 
virial BH mass is tens of times larger than the \msig expected value, leading SDSS J1241+2602 to be an outlier in the \msig 
space with confidence level higher than $5\sigma$. Therefore, the unobscured BLRs are preferred in the Type-1.9 AGN SDSS J1241+2602. 
The results indicate that it is necessary to check whether unobscured central BLRs are common in Type-1.9 AGN, when to test the 
AGN unified model of AGN by properties of Type-1.9 AGN.  
\end{abstract}


\section{Introduction}

	Both broad emission lines from central broad emission line regions (BLRs) and narrow emission lines from extended 
narrow emission line regions (NLRs) are fundamental spectroscopic characteristics in optical band of Type-1 AGN (broad emission line 
AGN) \citep{om86, sm00, oy15}. Meanwhile, strong narrow emission lines from central NLRs but no apparent broad emission lines are 
fundamental spectroscopic characteristics of Type-2 AGN (narrow emission line AGN). The well-known AGN unified model 
\citep{an93, nh15, bb18, kw21, zh22a} has been hypothesized to explain the different spectroscopic phenomena between Type-1 AGN and 
Type-2 AGN, after mainly considering severe obscurations on central BLRs in Type-2 AGN. Under the framework of the AGN unified model, 
Type-2 AGN and Type-1 AGN have intrinsically similar fundamental structures of central accretion disks around black holes (BHs), BLRs, 
dust torus and NLRs, but Type-2 AGN have central accretion disks around BHs and BLRs heavily obscured by central dust 
torus due to orientation with respect to the line of sight. The AGN unified model has been strongly supported by clearly 
detected polarized broad emission lines and/or clearly detected broad infrared emission lines in some Type-2 AGN \citep{tr03, sg18, mb20}.

	However, even considering different properties of both central dust torus and central BH accreting process expected 
properties, some challenges to the AGN unified model have been reported. \citet{fb02} have discussed probably different evolutionary 
patterns in Type-1 and Type-2 AGN. \citet{vk14} have reported different neighbours around Type-1 and Type-2 AGN. \citet{zy19} have 
reported lower host galaxy stellar masses in X-ray selected Type-1 AGN than Type-2 AGN. \citet{bg20} have shown significantly different 
properties of UV/optical and mid-infrared colour distributions of different AGN types. More recently, \citet{zh22b} have shown that 
direct measurements of stellar velocity dispersion can lead to statistically larger stellar velocity dispersions in Type-1 AGN than 
in Type-2 AGN, with a confidence level higher than 10$\sigma$, even after considering the necessary effects of different redshift 
and different physical properties related to central BH accreting processes in AGN. As discussed in \citet{nh15}, the AGN unified 
model has been successfully applied to explain different observed spectroscopic features between Type-1 and Type-2 AGN in many 
different ways, however, the AGN family with many other features considering the reported challenges to the AGN unified model are 
far from homogeneous.

	Besides Type-1 AGN and Type-2 AGN expected by the AGN unified model, there is a special kind of optically selected AGN, 
Type-1.9 AGN (firstly discussed in \citet{os81}), which have apparent broad H$\alpha$ but no apparent broad H$\beta$. Commonly, 
disappearance of broad H$\beta$ (or quite large broad Balmer decrements, large flux ratio of broad H$\alpha$ to broad H$\beta$) in 
Type-1.9 AGN are mainly attributed heavily obscured central BLRs, and can be applied to test the AGN unified model. 
However, as discussed in \citet{kk81, cp81, gr90}, BLRs modeled with relatively low optical depths and low ionization parameters can 
reproduce large broad Balmer decrements in Type-1.9 AGN, indicating there are rare Type-1.9 AGN of which central BLRs with large 
broad Balmer decrements are intrinsic but not due to serious obscuration. \citet{bx03} have discussed that the H1320+551 (a Type-1.9 
AGN) with no apparent broad H$\beta$ but apparent and strong broad H$\alpha$ is not consistent with being an obscured Type-1 AGN, 
through its unabsorbed X-ray properties. More recently, \citet{hm17} have discussed that Type-1.9 AGN and Type-2 AGN have different 
different variability properties in the UV and X-ray domains, indicating pure obscurations on central regions should be disfavored 
to explain different features between Type-1.9 AGN and Type-1/2 AGN. Here, in the manuscript, a Type-1.9 AGN, SDSS J124131.46+260233.57 
(=SDSS J1241+2602), is interestingly and firstly reported on its unobscured central BLRs with strong evidence from optical 
spectroscopic results.

	The manuscript is organized as follows. Section 2 shows the main hypotheses. Section 3 presents the spectroscopic results 
of \obj~ at redshift 0.0159. Section 4 describes our necessary discussions. Section 5 gives our final conclusions. And we have 
adopted the cosmological parameters of $H_{0}=70{\rm km\cdot s}^{-1}{\rm Mpc}^{-1}$, $\Omega_{\Lambda}=0.7$ and $\Omega_{\rm m}=0.3$.

\begin{figure*}
\centering\includegraphics[width = 18cm,height=13cm]{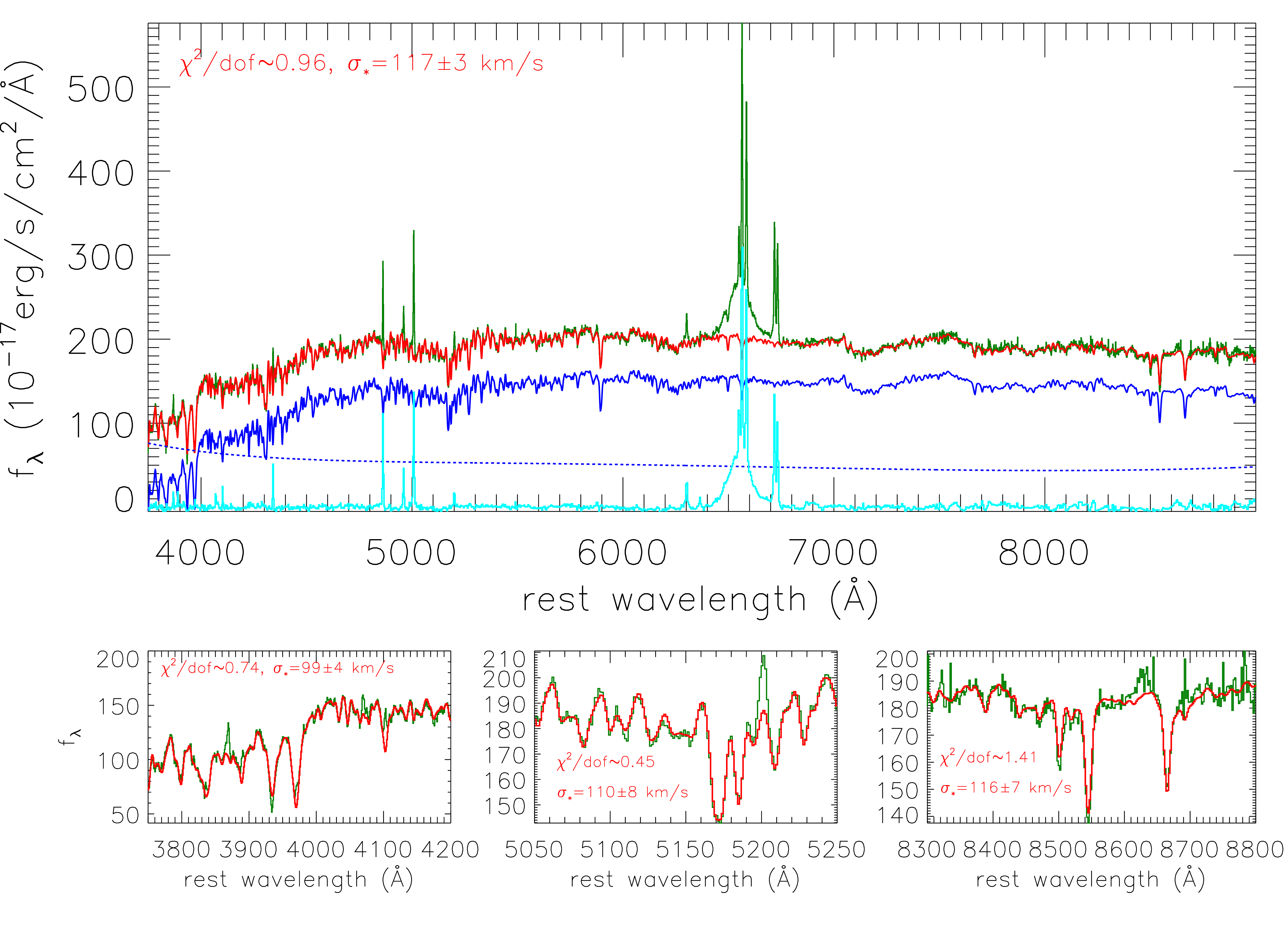}
\caption{Top panel shows the SSP method determined the best descriptions (solid red line) to the SDSS spectrum (solid dark green 
line) with emission lines being masked out. In top panel, solid blue line and dashed blue line show the determined host galaxy 
contributions and the determined AGN continuum emissions, respectively, solid cyan line shows the line spectrum calculated by the 
SDSS spectrum minus the sum of host galaxy contributions and AGN continuum emissions. Bottom panels show the best fitting results 
(solid red line) to absorption features (solid dark green line) of the Ca~{\sc ii} H+K (left panel), the Mg~{\sc i} (middle panel), 
the Ca~T (right panel). In each panel, the determined $\chi^2/dof$ and stellar velocity dispersion are marked in red characters.
}
\label{spec}
\end{figure*}

\begin{figure*}
\centering\includegraphics[width = 18cm,height=6cm]{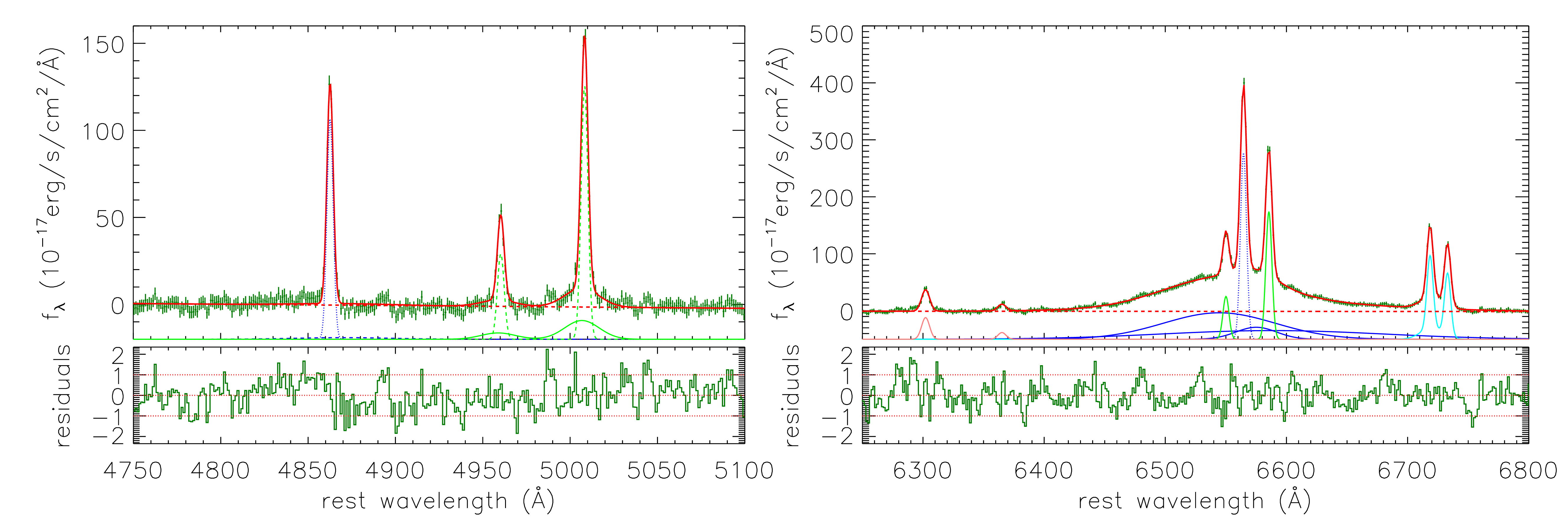}
\caption{Top panels show the best fitting results (solid red line) to the emission lines in the line spectrum (solid dark green 
line), and bottom panels show the corresponding residuals calculated by the line spectrum minus the best fitting results and then 
divided by uncertainties of the SDSS spectrum. In top left panel, dotted blue line shows the narrow H$\beta$, dotted and solid 
green lines show the core components and blue-shifted wings in [O~{\sc iii}] doublet. In top right panel, dotted blue line shows 
the narrow H$\alpha$, solid blue lines show the determined three broad Gaussian components in broad H$\alpha$, solid line in green, 
in pink and in cyan show the [N~{\sc ii}] doublet, the [O~{\sc i}] doublet and the [S~{\sc ii}] doublet. In each top panel, dashed 
red line shows the baseline $f_\lambda=0$. In each bottom panel, horizontal dashed lines show residuals=$0~,\pm1$, respectively.
}
\label{line}
\end{figure*}

\section{Main Hypotheses}

	In order to test heavily obscured central BLRs in a Type-1.9 AGN, properties of virial BH mass can be applied 
as follows. 

	Accepted the virialization assumption to central BLRs as discussed in \citet{ve02, pe04, sh11}, virial BH mass of a 
broad line AGN can be conveniently estimated by 
\begin{equation}
\frac{M_{BH}}{\rm 1.2131\times10^{-3}M_\odot}=5.5\times\frac{\frac{R_{BLRs}}{\rm 100light-days}\times
	(\frac{\sigma}{\rm 1000km/s})^2}{G}
\end{equation}, 
with $G=6.672\times10^{-11}{\rm Nm^2/kg^2}$ as the gravitational constant and $R_{BLRs}$ in units of 100light-days as distance 
of BLRs to central BH and $\sigma$ in units of km/s as line width (second moment) of broad emission lines to trace rotating 
velocities of broad line emission clouds in BLRs. The factor 5.5 is the virial factor, discussed in \citet{On04, wt10, gr11, pa12, 
wy15}. And, the $R_{BLRs}$ can be simply estimated through continuum luminosity by the improved empirical relation in \citet{bd13} 
after necessary corrections of host galaxy contaminations. Moreover, considering strong linear correlation between continuum 
luminosity and broad H$\alpha$ luminosity as discussed in \citet{gh05, mt22}, virial BH mass of a broad line AGN can be estimated 
by line width (second moment, $\sigma_{H\alpha}$) and line luminosity ($L_{H\alpha}$) of broad H$\alpha$
\begin{equation}
M_{BH}=15.6\times10^6(\frac{L_{H\alpha}}{\rm 10^{42}erg/s})^{0.55}
	(\frac{\sigma_{H\alpha}}{\rm 1000km/s})^{2.06}{\rm M_\odot}
\end{equation}
in order to ignore effects of uncertainties of measured continuum luminosities in broad line AGN with strong host galaxy 
contributions. Here, the second moment rather than the full width of maximum (FWHM) of broad H$\alpha$ is applied, mainly due 
to few effects of sharp features around peak of broad line profile on calculated second moment.

	Meanwhile, through measured stellar velocity dispersions of host galaxy stellar bulges, the \msig relation discussed 
in \citet{fm00, ge00, kh13, bb17, bt21} can also be conveniently applied to estimate central BH mass in both quiescent galaxies 
and active galaxies, without effects of obscurations on central BLRs. If there were intrinsically serious obscurations on central 
BLRs leading to disappearance of broad H$\beta$ in a Type-1.9 AGN, virial BH mass of the Type-1.9 AGN through properties of 
observed broad H$\alpha$ should be significantly smaller than the \msig relation expected value, which is the main point to test 
obscured/unobscured BLRs in a Type-1.9 AGN. In the next Section, stellar velocity dispersion of \obj~ can be measured through 
apparent absorption features, leading to measured \msig relation determined BH mass of \obj. Therefore, in \obj, interesting 
results are reported and discussed in the following sections on properties of \msig relation determined BH mass and on properties 
of virial BH mass with and without considerations of obscurations on central BLRs.

\section{Spectroscopic results of the Type-1.9 AGN \obj}

	\obj~ has its SDSS spectrum (plate-mjd-fiberid=2660-54504-0446) with signal-to-noise about 56 shown in Fig.~\ref{spec} 
with apparent broad H$\alpha$ and apparent stellar absorption features. In order to measure the emission lines, the commonly 
accepted SSP (Simple Stellar Population) method is applied to determine host galaxy contributions. More detailed descriptions 
on the SSP method can be found in \citet{bc03, ka03, cm05, cm17}. The SSP method has been applied in our previous papers 
\citet{zh21m, zh21a, zh21b, zh22a, zh22b}. Here, we show simple descriptions on SSP method as follows. The 39 simple stellar 
population templates from \citet{bc03, ka03} have been exploited, which can be used to describe the characteristics of 
almost all the SDSS galaxies. Meanwhile, there is an additional 5th-order polynomial component applied to describe intrinsic 
AGN continuum emissions. Here, as shown properties of the composite spectrum of SDSS quasars in \citet{vr01}, AGN continuum 
emissions can be fitted by two power laws with a break at 5000\AA, indicating a simple power law component not preferred. 
And moreover, higher-order polynomial functions are also checked, leading to none variability of the following determined 
$\chi^2/dof$ with $\chi^2$ as the summed squared residuals and $dof$ as degrees of freedom. Therefore, 
a 5th-order polynomial component is preferred in the manuscript. When the SSP method is applied, optical narrow emission lines 
are masked out by the full width at zero intensity (FWZI) about 450${\rm km~s^{-1}}$, and the spectrum with rest wavelength 
range from 6250 to 6750\AA~ are also masked out due to the strongly broad H$\alpha$. Then, through the Levenberg-Marquardt 
least-squares minimization technique (the known MPFIT package), the SDSS spectrum with emission lines being masked out can be 
described and shown in Fig.~\ref{spec} with corresponding $\chi^2/dof\sim0.96$ and with the determined stellar velocity 
dispersion about $117\pm3{\rm km/s}$ and with the determined continuum luminosity at rest wavelength 5100\AA~ (from the 
determine 5th-order polynomial component) about $(1.09\pm0.09)\times10^{42}{\rm erg/s}$.

	Before proceeding further, one point is noted. The SDSS pipeline reported stellar velocity dispersion is about 
$144\pm3{\rm km/s}$ in \obj, without considering AGN continuum emissions in the pipeline. If the 5th-order polynomial component 
is not considered, the similar procedure applied to describe the SDSS spectrum with only narrow emission lines being masked 
out can lead the determined stellar velocity dispersion to be $146\pm3{\rm km/s}$ consistent with SDSS pipeline reported 
value. However, considering the apparent broad H$\alpha$ in \obj, the 5th-order polynomial component should be preferred.

	Moreover, the Ca~{\sc ii} triplet (Ca~T) from 8300\AA~ to 8800\AA, the Ca~{\sc ii} H+K absorption features from 3750\AA~ 
to 4200\AA~ and the Mg~{\sc i} absorption features from 5050\AA~ to 5250\AA are applied to re-measure the stellar velocity 
dispersions of \obj, through the same SSP method discussed above to describe the whole SDSS spectrum. The best fitting results 
are shown in the bottom panels of Fig.~\ref{spec} with the determined stellar velocity dispersions in units of ${\rm km/s}$ about 
$116\pm7$, $99\pm4$ and $110\pm8$ through the Ca~T, the Ca~{\sc ii} H+K and the Mg~{\sc i} absorption features, respectively. 
Therefore, in the manuscript, the mean value $\sigma_\star=110\pm12{\rm km/s}$ is accepted as the stellar velocity dispersion of \obj.

	After subtractions of the host galaxy contributions and the AGN continuum emissions, emission lines in the line spectrum 
can be measured, similar as what we have previously done in \citet{zh21a, zh21b, zh22a, zh22b, zh22c}. For the emission lines 
within the rest wavelength range from 4750\AA~ to 5100\AA, there is one Gaussian function applied to describe narrow H$\beta$, 
four Gaussian functions applied to describe [O~{\sc iii}]$\lambda4959,5007$\AA~ doublet (two for the core components, and 
two for the blue-shifted wings). When the Gaussian functions above are applied, only two criteria are accepted. First, each 
Gaussian component has line intensity not smaller than zero. Second, the core (blue-shifted) components of the [O~{\sc iii}] 
doublet have the same redshift and the same line width, and have the flux ratio to be fixed to the theoretical value 3. Then, 
through the Levenberg-Marquardt least-squares minimization technique, the best fitting results to the emission lines and the 
corresponding residuals (calculated by the line spectrum minus the best fitting results then divided by the uncertainties of the 
SDSS spectrum) are shown in left panels of Fig.~\ref{line} with $\chi_0^2/dof_0=209.8/344\sim0.61$. Besides the discussed Gaussian 
components, one additional broad Gaussian function was tried to be applied to describe probable broad H$\beta$, however, the 
fitting procedure led the additional broad component to have the determined line flux and line width smaller than their 
corresponding determined uncertainties. Therefore, it is not necessary to consider broad Gaussian components to describe the broad 
H$\beta$ in \obj.

	Meanwhile, emission lines within the rest wavelength range from 6250\AA~ to 6800\AA~ can also be measured by multiple 
Gaussian functions. There is one Gaussian function applied to describe the narrow H$\alpha$, three broad Gaussian functions 
applied to describe the broad H$\alpha$, two Gaussian functions applied to describe the [N~{\sc ii}] doublet, two Gaussian 
functions applied to describe the [O~{\sc i}] doublet, four Gaussian functions applied to describe the [S~{\sc ii}] doublet 
(two for the core components and two for the shifted wings). When the Gaussian functions above are applied, only three criteria 
are accepted. First, each Gaussian component has line intensity not smaller than zero. Second, the components of the [N~{\sc ii}] 
(the [O~{\sc i}], the [S~{\sc ii}]) doublet have the same redshift and the same line width, and the [N~{\sc ii}] doublet have 
the flux ratio to be fixed to the theoretical value 3. Third, the components in the narrow H$\alpha$ and in the narrow H$\beta$ 
have the same redshift and the same line width. Then, through the Levenberg-Marquardt least-squares minimization technique, 
the best fitting results and the corresponding residuals are shown in right panels of Fig.~\ref{line} with 
$\chi^2/dof=160.6/398\sim0.41$. Besides the discussed Gaussian components above, additional Gaussian functions were tried to 
be applied to describe probable blue/red-shifted wings of the [O~{\sc i}] and the [N~{\sc ii}] doublets, however, the fitting 
procedure led the determined line fluxes of the additional Gaussian components to be smaller than their corresponding determined 
uncertainties. Therefore, it is not necessary to consider additional blue/red-shifted wings in the [O~{\sc i}] and the [N~{\sc ii}] 
doublets in \obj.

	Before proceeding further, one point is noted. If different numbers (not three) of broad Gaussian functions were applied 
to describe the broad H$\alpha$ of \obj, whether were there different results on line profiles? In order to test effects of 
applications of different numbers of broad Gaussian functions, the F-test technique is applied, similar as what we have recently 
done in \citet{zh22c}. For one, two and four broad Gaussian functions applied to describe broad H$\alpha$ of \obj, the corresponding 
$\chi^2/dof$ are about $\chi_1^2/dof_1=238.8/404\sim0.59$, $\chi_2^2/dof_2=224.2/401\sim0.56$ and $\chi_4^2/dof_4=160.6/395\sim0.406$, 
respectively. Then, based on the different $\chi^2$ and $dof$ for different model functions, the F-test technique can be applied 
to confirm the confidence level higher than 5$\sigma$ to support that three broad Gaussian functions are referred to describe 
the broad H$\alpha$ than one or two broad Gaussian functions, and the F-test technique can be applied to confirm the probability 
only about $10^{-5}$ to support that the four broad Gaussian functions are preferred to describe the broad H$\alpha$ than the 
three broad Gaussian functions. Therefore, in the manuscript, three broad Gaussian functions are applied to describe the broad 
H$\alpha$ of \obj.

\begin{table}
\caption{Line parameters of each Gaussian emission component}
\begin{tabular}{llll}
\hline\hline
line & $\lambda_0$ & $\sigma$ & flux \\
\hline\hline
\multirow{3}{*}{Broad H$\alpha$}  & 6545.8$\pm$2.9 & 45.6$\pm$2.5 & 5342$\pm$832 \\
	& 6575.3$\pm$1.9  & 16.2$\pm$2.1  & 839$\pm$196 \\
	& 6582.1$\pm$12.4 & 95.9$\pm$13.7 & 3787$\pm$557 \\
\hline
Narrow H$\alpha$ & 6564.6$\pm$0.1 & 2.5$\pm$0.1 &  2072$\pm$38 \\  
\hline
Narrow H$\beta$ & 4862.6$\pm$0.1 & 1.9$\pm$0.1 &  594$\pm$17 \\
\hline
\multirow{2}{*}{[O~{\sc iii}]$\lambda5007$\AA} & 5008.3$\pm$0.1 & 1.9$\pm$0.1 & 710$\pm$25 \\
	& 5006.7$\pm$1.1 & 9.1$\pm$1.3 & 270$\pm$34 \\
\hline
[O~{\sc i}]$\lambda6300$\AA & 6302.1$\pm$0.2 & 3.3$\pm$0.2 &  317$\pm$18 \\
\hline
[O~{\sc i}]$\lambda6363$\AA & 6365.2$\pm$0.2 & 3.4$\pm$0.7  & 105$\pm$18 \\
\hline
[N~{\sc ii}]$\lambda6583$\AA & 6585.5$\pm$0.1 & 2.5$\pm$0.1 & 1406$\pm$32 \\
\hline
\multirow{2}{*}{[S~{\sc ii}]$\lambda6716$\AA} & 6718.5$\pm$0.1 & 2.5$\pm$0.1 &  809$\pm$54 \\
	& 6719.8$\pm$1.7 & 9.2$\pm$1.9 & 446$\pm$81 \\
	\hline
\multirow{2}{*}{[S~{\sc ii}]$\lambda6731$\AA} & 6733.1$\pm$0.1 & 2.6$\pm$0.1 & 707$\pm$46 \\
	& 6734.4$\pm$1.9 & 9.2$\pm$1.9 & 3$\pm$2 \\
\hline\hline
\end{tabular}\\
\tablecomments{The first column shows which line is measured. The Second, third, fourth columns show the measured 
line parameters: the center wavelength $\lambda_0$ in units of \AA, the line width (second moment) $\sigma$ 
in units of \AA~ and the line flux in units of ${\rm 10^{-17}~erg/s/cm^2}$. \\	
For broad H$\alpha$, there are three Gaussian components. For [O~{\sc iii}]$\lambda5007$\AA~ and each 
[S~{\sc ii}] emission line, there are two components: the core and the shifted-wing related extended components.}
\end{table}

	Based on the measured line parameters listed in Table~1, the observed line luminosities are 
$(3.39\pm0.11)\times10^{39}{\rm erg/s}$ and $(1.18\pm0.02)\times10^{40}{\rm erg/s}$ for the narrow H$\beta$ and the narrow 
H$\alpha$, respectively, leading to normal flux ratio 3.5 of the narrow H$\alpha$ to the narrow H$\beta$. The line width is 
about $114\pm2{\rm km/s}$ for the narrow emission lines, consistent with the measured stellar velocity dispersion. The 
line luminosity and the line width (second moment) of the broad H$\alpha$ (summed the three broad Gaussian components) are 
about $(5.69\pm0.91)\times10^{40}{\rm erg/s}$ and about $3150\pm420{\rm km/s}$, respectively. The uncertainties above are 
determined through the determined uncertainties of the Gaussian emission components. The results can be applied to confirm 
that \obj~ is a Type-1.9 AGN with apparent broad H$\alpha$ but no broad H$\beta$.

\begin{figure*}
\centering\includegraphics[width = 18cm,height=12cm]{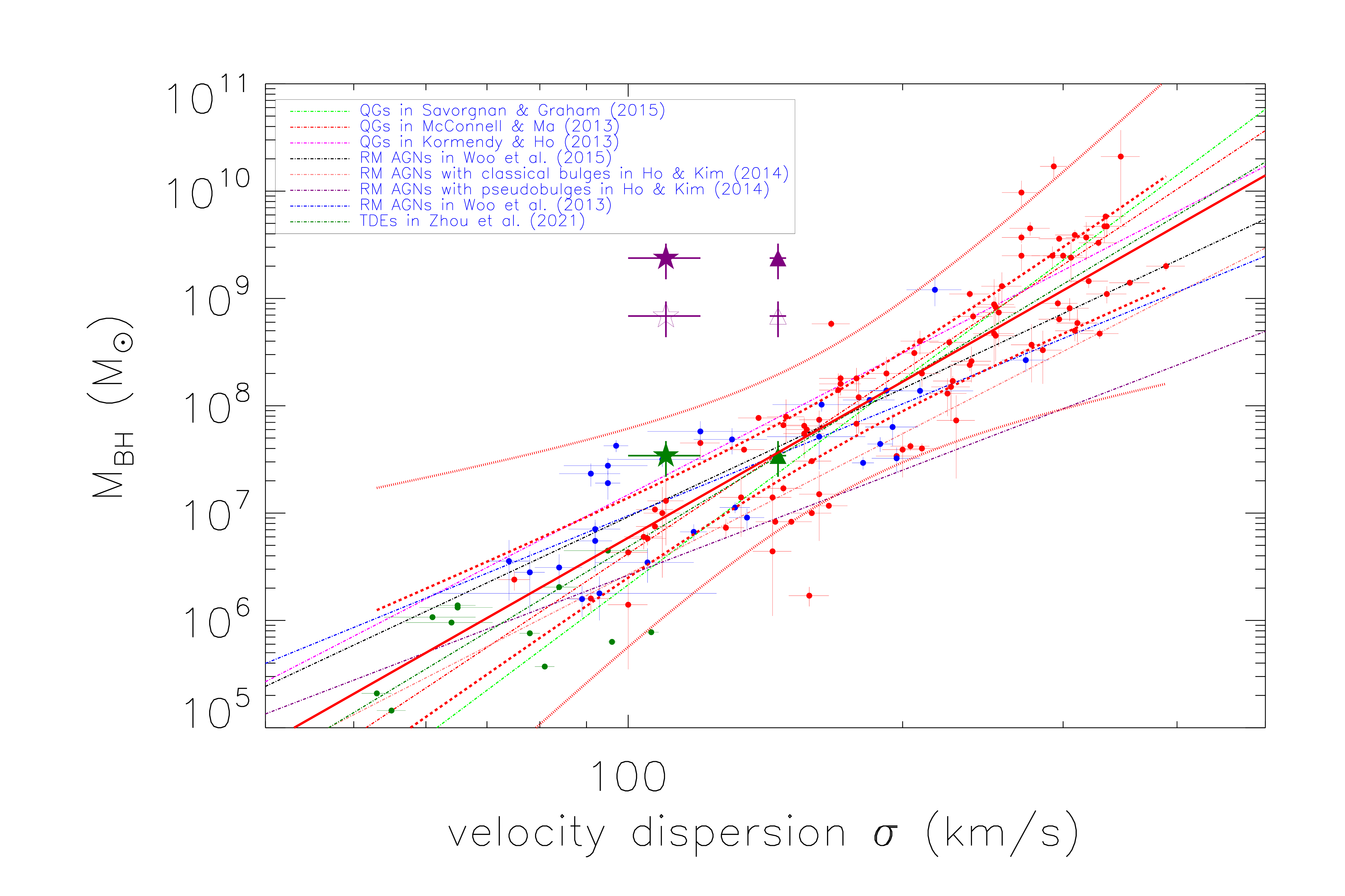}
\caption{On the correlation between stellar velocity dispersion measured through the absorption features and the virial BH 
mass of \obj. Solid five-point-star in dark green shows the virial BH mass of \obj~ determined by properties of the observed 
broad H$\alpha$ without considerations of any obscurations on central BLRs. Solid and open five-point-star in purple show the 
virial BH mass of \obj~ determined by properties of the reddening corrected broad H$\alpha$ with $E(B-V)\sim3.4$ (3$\sigma$ 
confidence level for upper limit of broad H$\beta$) and $E(B-V)\sim2.6$ (5$\sigma$ confidence level for upper limit of broad 
H$\beta$), respectively. Dot-dashed lines in green, in red, in magenta, in black, in pink, in purple, in blue and in dark green 
represent the \msig relations through the quiescent galaxies in \citet{sg15}, in \citet{mm13}, in \citet{kh13}, and through the 
RM AGNs in \citet{wy15}, the RM AGNs with classical bulges in \citet{hk14}, the RM AGNs with pseudobulges in \citet{hk14} and 
the RM AGNs in \citet{ws13}, and through the TDEs in \citet{zl21}, respectively. Solid circles in red, in blue and in cyan show 
the values for the 89 quiescent galaxies from \citet{sg15}, the 29 RM AGNs from \citet{wy15} and the 12 TDEs from \citet{zl21}, 
respectively. Thick solid red line shows the best fitting results to all the objects, and thick dashed and dotted red lines show 
the corresponding $3\sigma$ and $5\sigma$ confidence bands to the best fitting results. If accepted the stellar 
velocity dispersion about $146\pm3$km/s of \obj~ without considerations of AGN continuum emissions, solid triangle in dark green 
shows the virial BH mass of \obj~ determined by properties of the observed broad H$\alpha$ without considerations of any 
obscurations on central BLRs, solid and open triangles in purple show the virial BH mass of \obj~ determined by properties of 
the reddening corrected broad H$\alpha$ with $E(B-V)\sim3.4$ (3$\sigma$ confidence level for upper limit of broad H$\beta$) and 
$E(B-V)\sim2.6$ (5$\sigma$ confidence level for upper limit of broad H$\beta$), respectively.}
\label{msig}
\end{figure*}

\begin{figure*}
\centering\includegraphics[width = 18cm,height=5cm]{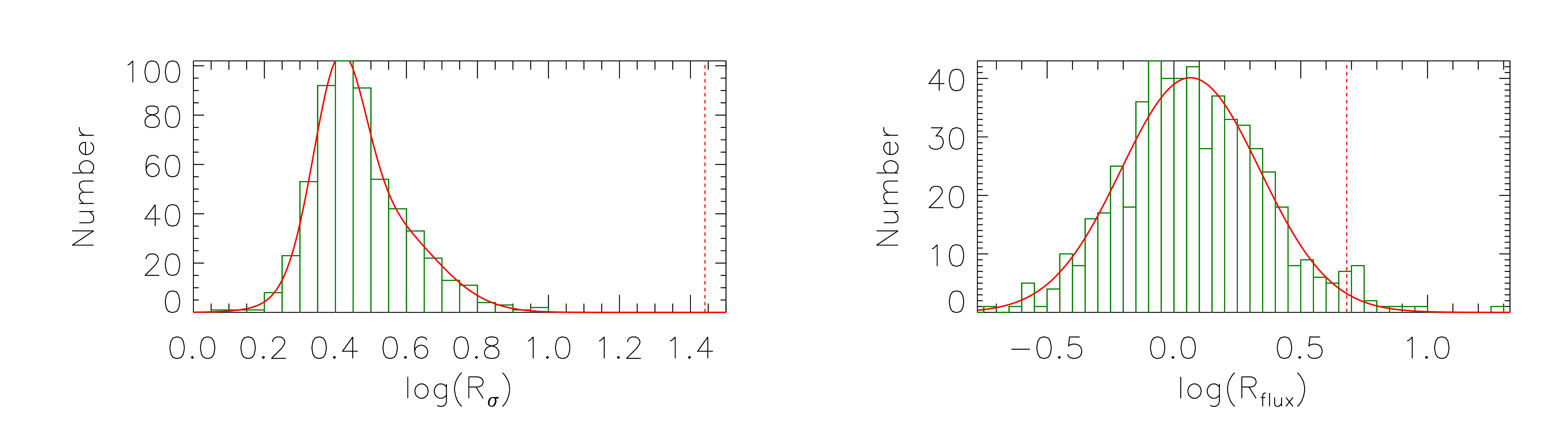}
\caption{Distributions of $\log(R_{\sigma})$ and $\log(R_{flux})$ of the shifted-wing related broad component to the core 
component of [O~{\sc iii}]$\lambda5007$\AA~ of the 557 blue quasars discussed in \citet{zh21a}. In each panel, solid red line 
shows the multiple-Gaussian-function description to the distribution, and vertical dashed red line marks position of \obj.}
\label{dis}
\end{figure*}

\begin{figure}
\centering\includegraphics[width = 8cm,height=5cm]{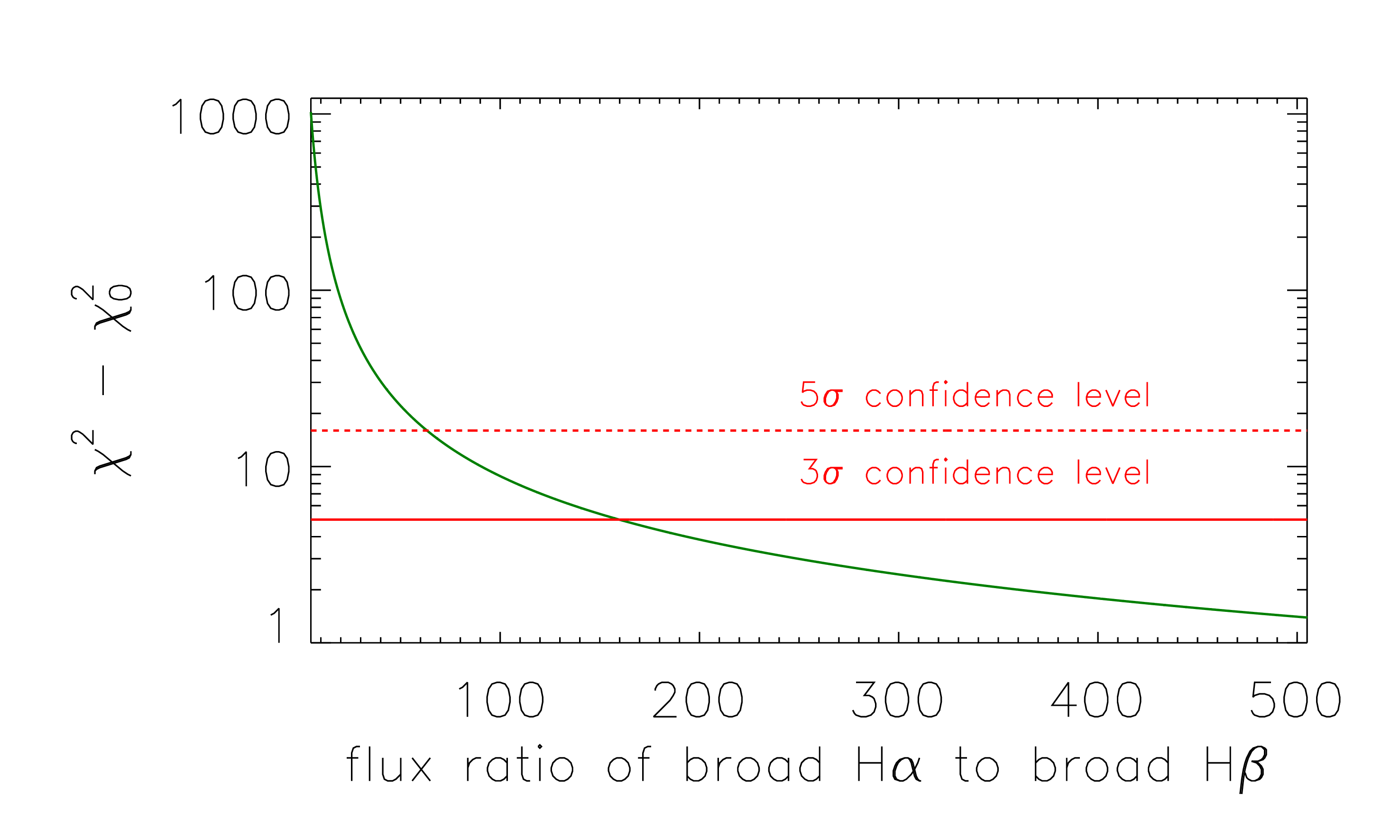}
\caption{Dependence of $\chi^2-\chi^2_0$ on $BD_b$. Horizontal solid and dotted red lines show the $3\sigma$ and $5\sigma$ 
confidence levels, respectively.}
\label{fake}
\end{figure}

\section{Unobscured central BLRs in the Type-1.9 AGN \obj}

	Based on the calculated line width (second moment) and line luminosity of the broad H$\alpha$ without considering any 
obscurations on central BLRs, the virial BH mass in \obj~ can be estimated as $(3.43\pm1.25)\times10^7{\rm M_\odot}$ through 
equation (2) above. The uncertainty of virial BH mass is estimated by uncertainties of the line width and the line luminosity 
of the broad H$\alpha$. Meanwhile, similar virial BH mass about $2.5\times10^7{\rm M_\odot}$ in \obj~ have been reported in 
\citet{ld19, ms22}. It is interesting that the virial BH mass is consistent with the \msig relation expected BH mass. The 
dependence of virial BH mass on stellar velocity dispersion in \obj~ is shown in Fig.~\ref{msig} with previous reported results 
in the literature for quiescent galaxies, active galaxies and also in Tidal disruption events (TDEs), etc.. The corresponding 
reference can be found in the caption of Fig.~\ref{msig}.

	Before proceeding further, it is interesting to check whether a broad component related to shifted wings of narrow Balmer 
lines can also lead to so large virial BH mass. If accepted that the broad components shown in Fig.~\ref{line} and listed in Table~1 
are components related to shifted wing of narrow H$\alpha$, the line width ratio $R_\sigma$ and the line flux ratio $R_{flux}$ 
are about 27.6 and 4.8 of the shifted-wing related extended component to the core component of the narrow H$\alpha$ in \obj. If 
simply assumed similar properties of the road components related to the shifted wings of the narrow H$\alpha$ as those discussed 
in blue-shifted components in [O~{\sc iii}] lines of the 557 blue quasars in \citet{zh21a} with distributions of line width ratio 
$R_\sigma$ and flux ratio $R_{flux}$ of the broad component to the core component of [O~{\sc iii}]$\lambda5007$\AA~ shown in 
Fig.~\ref{dis}, probability about $1.54\times10^{-10}$ for $\log(R_\sigma)>\log(27.6)$ and probability about 1.21\% for 
$\log(R_{flux})>\log(4.8)$. Therefore, the probability is only about $1.54\times10^{-10}\times1.21\%\sim1.86\times10^{-12}$ for a 
shifted-wing related broad component in \obj, leading to the similar virial BH mass. In other words, confidence level is higher 
than 5$\sigma$ to support that the broad components in Balmer lines are not components related to shifted wings of narrow Balmer 
lines but really related to central BLRs in \obj.

	However, if accepted the disappearance of broad H$\beta$ was due to serious obscurations on central BLRs in \obj, upper 
limits of line flux of the broad H$\beta$ can be simply estimated through F-test technique as follows, and then different virial 
BH mass can be estimated. Assumed the obscured broad H$\beta$ overwhelmed in spectral noises have the same line profile as that 
($[\lambda(H\alpha_b)$,~~~$f_\lambda(H\alpha_b)$]) of broad H$\alpha$ described by three Gaussian functions but have total line 
flux determined by the line flux of broad H$\alpha$ divided by the expected broad Balmer decrement $BD_b$, then the intrinsic 
but obscured broad H$\beta$ can be described as
\begin{equation}
[\lambda(H\beta_b),~~~~f_\lambda(H\beta_b)]~=~[\lambda(H\alpha_b)\times
	\frac{4862.68\textsc{\AA}}{6564.61\textsc{\AA}},~~~~\frac{f_\lambda(H\alpha_b)}{BD_b}]
\end{equation}
For the observed line spectrum shown in top left panel of Fig.~\ref{line}, contributions of obscured broad H$\beta$ can be 
considered, and new line spectrum can be created as
\begin{equation}
[\lambda(H\beta_b),~~~~f_\lambda]~=~[\lambda(H\beta_b),~~~~f_{\lambda,~obs}~+~f_\lambda(H\beta_b)]
\end{equation}
Then, for a series of 1000 new line spectrum including different contributions of obscured broad H$\beta$ with different 
values of $BD_b$ (larger than 3), $\chi^2$ values can be calculated 
\begin{equation}
	\chi^2_i~=~\sum(\frac{f_\lambda~-~Y_{fit}}{y_{err}})^2~~~~~~(i=1,~\dots,~1000)
\end{equation}
with $Y_{fit}$ as the shown best fitting results in top left panel of Fig.~\ref{line} and $y_{err}$ as uncertainties of 
SDSS spectrum. Then, through the F-test statistical technique applied with $1$ (only one additional parameter of $BD_b$) 
and $343$ ($dof_0-1$) as number of dofs of the F-distribution numerator and denominator, the calculated $3\sigma$ and 
$5\sigma$ confidence levels $CL~\sim~(\frac{\chi^2-\chi^2_0}{1})/(\frac{\chi^2}{343})$ for strong broad H$\beta$ can lead 
$\chi^2-\chi^2_0$ to be 5 and 16, respectively. Here, $\chi^2_0$ and $dof_0$ are for the best fitting results shown in top 
left panel of Fig.~\ref{line}.

	The dependence of $\chi^2-\chi^2_0$ on $BD_b$ is shown in Fig.~\ref{fake}, leading to $BD_b\sim160$ and $BD_b\sim64$ 
for the obscured broad H$\beta$ with $3\sigma$ and $5\sigma$ confidence levels, leading $E(B-V)$ to be around 3.4 and 2.6, 
assumed the intrinsic flux ratio 3.1 of broad H$\alpha$ to broad H$\beta$. Therefore, considering serious obscurations on central 
BLRs with $E(B-V)\sim3.4$ ($E(B-V)\sim2.6$), the reddening corrected line luminosity of the broad H$\alpha$ should be about 1605 
(283) times higher than the value from the observed broad H$\alpha$, leading the reddening corrected virial BH mass to be about 
58 (22) times higher than the value from properties of the observed broad H$\alpha$. The corrected virial BH masses are shown as 
solid/open five-point-stars in purple in Fig.~\ref{msig}, quite larger than the \msig relation expected values.

	In order to show more clear results in Fig.~\ref{msig}, the 89 quiescent galaxies from \citet{sg15} and the 29 
reverberation mapped (RM) AGN from \citet{wy15} and the 12 tidal disruption events (TDEs) from \citet{zl21} are considered 
to draw the linear correlation between stellar velocity dispersion and BH mass
\begin{equation}
\log(\frac{M_{BH}}{\rm M_\odot})=(-2.89\pm0.49)+(4.83\pm0.22)\times\log(\frac{\sigma_\star}{\rm km/s}) 
\end{equation}
through the Least Trimmed Squares robust technique \citep{cm13}. And then the $3\sigma$ and $5\sigma$ confidence bands to the 
linear correlation are determined and shown in Fig.~\ref{msig}. Therefore, the reddening corrected viral BH mass should lead 
the \obj~ as an outlier with confidence levels higher than $5\sigma$.

	Before end of the section, one point is noted. As shown in Section 3, large stellar velocity dispersion $\sim146$km/s 
can be estimated in \obj, if not considering AGN continuum emissions apparently included in the SDSS spectrum. It is necessary to 
discuss whether the larger stellar velocity dispersions can affect our final results shown in Fig.~\ref{msig}. Here, we consider 
the question by the following two points. On the one hand, the F-test technique can be applied to confirm the confidence level 
higher than $5\sigma$ for the AGN continuum emissions described by the 5th-order polynomial function included in model functions 
to describe the SDSS spectrum of SDSS J1241+2602, based on the $\chi^2/dof=3272.56/3382\sim0.96$ and the 
$\chi^2/dof=3972.38/3387\sim1.17$ for the SSP method determined best descriptions to the SDSS spectrum with and without 
considerations of AGN continuum emissions. Therefore, the stellar velocity dispersion $\sim$110km/s is preferred in \obj, 
considering contributions of AGN continuum emissions to the SDSS spectrum. On the other hand, even without considerations of 
AGN continuum emissions, properties of SDSS J1241+2602 with the larger stellar velocity dispersion about $146\pm3$km/s are also 
shown in Fig.~\ref{msig} as triangles in different colors, to re-support that the reddening corrected viral BH mass can also lead 
the SDSS J1241+2602 to be an unique outlier in the space of viral BH mass versus stellar velocity dispersion. Therefore, different 
stellar velocity dispersions with and without considerations of AGN continuum emissions have few effects on our final conclusions.

	Considering the \msig relation (no effects from obscurations on central BLRs) expected BH mass, the heavily obscured  
central BLRs should be disfavoured in \obj, indicating unobscured BLRs in the Type-1.9 AGN \obj~ with apparent broad H$\alpha$ 
but no broad H$\beta$. Besides the \obj~ discussed in the manuscript, H1320+551 is the other individual Type-1.9 
AGN reported in the literature \citep{bx03} with unobscured BLRs. Unfortunately, there is no clear information of stellar velocity 
dispersion in H1320+551, leading to no further discussions on BH mass properties determined through different methods as discussed 
in the manuscript. But in the near future, it is interesting to check virial BH mass properties of a large sample of 
Type-1.9 AGN, to test whether unobscured BLRs are common in Type-1.9 AGN, and then to provide clues on evolution of different 
Types of AGN under the framework of the AGN unified model..

\section{Conclusions}

	Based on the measured stellar velocity dispersion through the absorption features in the Type-1.9 AGN \obj~ with apparent 
broad H$\alpha$ but no broad H$\beta$, the \msig relation expected BH mass is consistent with the virial BH mass through the 
observed broad H$\alpha$ without considering any obscurations on central BLRs. Meanwhile, if considering serious obscurations 
on central BLRs to explain the disappearance of broad H$\beta$ in the Type-1.9 AGN \obj, the reddening corrected broad H$\alpha$ 
line luminosity should lead \obj~ having the re-calculated reddening corrected virial BH mass to be an outlier in the \msig space 
with confidence level higher than $5\sigma$. Based on the properties of virial BH mass, the unobscured central BLRs is favoured 
in the Type-1.9 AGN \obj. The results indicate that obscured/unobscured BLRs of Type-1.9 AGN should be firstly discussed, 
when to test the AGN unified model by properties of Type-1.9 AGN.

\section*{Acknowledgements}
Zhang gratefully acknowledge the anonymous referee for giving us constructive comments and suggestions to greatly 
improve our paper.  
Zhang gratefully acknowledges the research funding support from GuangXi University and the kind funding support from 
NSFC-12173020 and NSFC-12373014. This research has made use of the data from the SDSS (\url{https://www.sdss.org/}) funded by 
the Alfred P. Sloan Foundation, the Participating Institutions, the National Science Foundation and the U.S. Department of Energy 
Office of Science. The research has made use of the MPFIT package \url{https://pages.physics.wisc.edu/~craigm/idl/cmpfit.html} 
to solve the least-squares problem through the Levenberg-Marquardt technique, and of the LTS\_LINEFIT package 
\url{https://www-astro.physics.ox.ac.uk/~cappellari/software/} to do linear fitting through Least Trimmed Squares robust technique.


\end{document}